\documentstyle[psfig]{mn}

\newif\ifAMStwofonts

\def\HII{H{\sevensize II}}

\ifoldfss
  \ifCUPmtlplainloaded \else
    \NewTextAlphabet{textbfit} {cmbxti10} {}
    \NewTextAlphabet{textbfss} {cmssbx10} {}
    \NewMathAlphabet{mathbfit} {cmbxti10} {} 
    \NewMathAlphabet{mathbfss} {cmssbx10} {} 
  \fi
  \ifAMStwofonts
    \ifCUPmtlplainloaded \else
      \NewSymbolFont{upmath} {eurm10}
      \NewSymbolFont{AMSa} {msam10}
      \NewMathSymbol{\upi}     {0}{upmath}{19}
      \NewMathSymbol{\umu}     {0}{upmath}{16}
      \NewMathSymbol{\upartial}{0}{upmath}{40}
      \NewMathSymbol{\leqslant}{3}{AMSa}{36}
      \NewMathSymbol{\geqslant}{3}{AMSa}{3E}

    \fi
  \fi
\fi 

\ifnfssone
  \newmathalphabet{\mathit}
  \addtoversion{normal}{\mathit}{cmr}{m}{it}
  \addtoversion{bold}{\mathit}{cmr}{bx}{it}
  \newmathalphabet{\mathbfit} 
  \addtoversion{normal}{\mathbfit}{cmr}{bx}{it}
  \addtoversion{bold}{\mathbfit}{cmr}{bx}{it}
  \newmathalphabet{\mathbfss} 
  \addtoversion{normal}{\mathbfss}{cmss}{bx}{n}
  \addtoversion{bold}{\mathbfss}{cmss}{bx}{n}
  \ifAMStwofonts
    \ifCUPmtlplainloaded \else
      \UseAMStwoboldmath
      \makeatletter
      \new@mathgroup\upmath@group
      \define@mathgroup\mv@normal\upmath@group{eur}{m}{n}
      \define@mathgroup\mv@bold\upmath@group{eur}{b}{n}
      \edef\UPM{\hexnumber\upmath@group}
      \new@mathgroup\amsa@group
      \define@mathgroup\mv@normal\amsa@group{msa}{m}{n}
      \define@mathgroup\mv@bold\amsa@group{msa}{m}{n}
      \edef\AMSa{\hexnumber\amsa@group}
      \makeatother
      \mathchardef\upi="0\UPM19
      \mathchardef\umu="0\UPM16
      \mathchardef\upartial="0\UPM40
      \mathchardef\leqslant="3\AMSa36
      \mathchardef\geqslant="3\AMSa3E
    \fi
  \fi
\fi 

\ifnfsstwo
 \newcommand{\hii}{\relax \ifmmode {\mbox H\,{\scshape ii}}\else H\,{\scshape ii}\fi}
\newcommand{\mi}{\relax \ifmmode {\mu{\mbox m}}\else $\mu$m\fi}
\newcommand{\ha}{\relax \ifmmode {\mbox H}\alpha\else H$\alpha$\fi}
\newcommand{\hb}{\relax \ifmmode {\mbox H}\beta\else H$\beta$\fi}
\newcommand{\degree}{\hbox{$^\circ$}}
\newcommand{\sii}{\relax \ifmmode {\mbox S\,{\scshape ii}}\else S\,{\scshape ii}\fi}
\newcommand{\siii}{\relax \ifmmode {\mbox S\,{\scshape iii}}\else S\,{\scshape iii}\fi}
\newcommand{\nii}{\relax \ifmmode {\mbox N\,{\scshape ii}}\else N\,{\scshape ii}\fi}
\newcommand{\oii}{\relax \ifmmode {\mbox O\,{\scshape ii}}\else O\,{\scshape ii}\fi}
\newcommand{\oiii}{\relax \ifmmode {\mbox O\,{\scshape iii}}\else O\,{\scshape iii}\fi}
 
 \newcommand{\rdostres}{\relax \ifmmode {\,\mbox{R}}_{\rm 23}\else \,\mbox{R}$_{\rm 23}$\fi}

  \DeclareMathAlphabet{\mathbfit}{OT1}{cmr}{bx}{it}
  \SetMathAlphabet\mathbfit{bold}{OT1}{cmr}{bx}{it}
  \DeclareMathAlphabet{\mathbfss}{OT1}{cmss}{bx}{n}
  \SetMathAlphabet\mathbfss{bold}{OT1}{cmss}{bx}{n}
  \ifAMStwofonts
    \ifCUPmtlplainloaded \else
      \DeclareSymbolFont{UPM}{U}{eur}{m}{n}
      \SetSymbolFont{UPM}{bold}{U}{eur}{b}{n}
      \DeclareSymbolFont{AMSa}{U}{msa}{m}{n}
      \DeclareMathSymbol{\upi}{0}{UPM}{"19}
      \DeclareMathSymbol{\umu}{0}{UPM}{"16}
      \DeclareMathSymbol{\upartial}{0}{UPM}{"40}
      \DeclareMathSymbol{\leqslant}{3}{AMSa}{"36}
     \DeclareMathSymbol{\geqslant}{3}{AMSa}{"3E}
    \fi
  \fi
\fi 

\ifCUPmtlplainloaded \else
  \ifAMStwofonts \else 
    \def\upi{\pi}
    \def\umu{\mu}
    \def\upartial{\partial}
  \fi
\fi

\title[A model for NGC~595]{A photoionization model of the spatial distribution of the optical and mid-IR
properties in NGC~595}

\author[E. P{\'e}rez-Montero et al.]
       {E. P{\'e}rez-Montero$^{1}$, M. Rela\~no$^2$, J. M. V\'\i lchez$^{1}$  \& A. Monreal-Ibero$^{1,3}$\\
$^{1}$ Instituto de Astrof\'\i sica de Andaluc\'\i a. CSIC. Apartado de correos 3004. 18080, Granada, Spain.\\
$^2$ Institute of Astronomy. University of Cambridge, Madingley Road, Cambridge, CB3 0HA, UK\\
$^3$ Astrophysikalisches Institut Postdam, An der Sternwarte 16, 14482 Postdam, Germany }
        
\date{Accepted 
      Received ;
      in original form November 2006}

\pagerange{\pageref{firstpage}--\pageref{lastpage}}
\pubyear{2006}

\begin{document}

\maketitle

\label{firstpage}

\begin{abstract}
We present a set of photoionization models that reproduce simultaneously the observed optical and mid-infrared spatial distribution 
of the \hii\ region NGC~595 in the disk of M33 using the code CLOUDY.  Both optical (PMAS-Integral Field Spectroscopy) 
and mid-infrared (8\,\mi\ and 24\,\mi\ bands from Spitzer) data provide enough spatial resolution to model in a novel approach 
the inner structure of the \hii\ region. We define a set of elliptical annular regions around the central ionizing
cluster with an uniformity in their observed properties and consider each annulus as an 
independent thin shell structure. For the first time our models fit the relative surface
brightness profiles in both the optical (H$\alpha$, [\oii], [\oiii]) and the mid-infrared emissions (8\,\mi\ and 24\,\mi), under the assumption
of a uniform metallicity (12+log(O/H) = 8.45; Esteban et al. 2009) and an age for 
the stellar cluster of 4.5\,Myr (Malumuth et al. 1996). Our models also reproduce the observed uniformity
of the R$_{23}$ parameter and the increase of the [\oii]/[\oiii] ratio due to the decrease of
the ionization parameter. The variation of the H$\alpha$ profile is explained in terms of the differences
of the occupied volume (the product of filling factor and total volume of the shell) in a
matter-bounded geometry, which also allows to reproduce the observed pattern of the extinction.  
The 8\,\mi/24\,\mi\ ratio is low (ranging between 0.04 and 0.4) because it is dominated
by the surviving of small dust grains in the \hii\ region, while the PAHs emit more weakly because
they cannot be formed in these thin \hii\ gas shells. The ratio is also well fitted in our
models by assuming a dust-to-gas ratio in each annulus compatible with the integrated estimate for the whole
\hii\ region after the 70\,\mi\, and 160\,\mi\, Spitzer observations.
\end{abstract}


\section{Introduction}


The approach to study the physical properties of \hii\ regions is changing nowadays with the
 availability of spectroscopic observations covering the whole face of these objects. 
 The observations are normally done with Integral Field Spectroscopy (IFS) instruments 
such as 
PMAS (\emph{Potsdam Multi Aperture Spectrophotometer}, S{\'a}nchez, 
Cardiel \& Verheijen et~al. 2007), VIMOS
(Le F{\`e}vre et al., 2003), and others.
These observations present a challenge for the classical 
photoionization models, which have normally been applied to explain long-slit observations 
typically centered at the location of the most intense knots within the regions. 
3D photoionization models allow the simulation of the thermal and ionization
structure in different geometries assuming different distributions for the
ionizing sources. Using these techniques Wood et al. (2004) studied the variation of the emission
line intensities in a stratified structure, with an increase of the temperature and
the ratios involving low excitation ions, like [\nii]/H$\alpha$ or [\sii]/H$\alpha$.
More recently, 3D photoionization models of \hii\ regions show that some of the physical parameters normally 
assumed to be constant are influenced by the configuration of the stars and gas within the 
region (Ercolano, Bastian \& Stasi{\'n}ska 2007). 
Nevertheless, the most important constraint to this approach is the assumed symmetry of
the gas configuration in the models. This approach can hardly be 
applied when comparing with IFS-based data, given the complex structure 
of most GHIIRs, mainly due to the interplay between the stars, gas and dust.

In a recent paper, Rela\~no et al. (2010) presented IFS observations of NGC~595, covering the 
major fraction of its surface. These authors created maps of the main emission lines observed
 within the 3650-6990~\AA\ spectral range and study the variations of the physical properties
across the surface of the region. The [\sii]$\lambda$6717/[\sii]$\lambda$6731 emission line
ratio map, tracing the electron density, does not show any relevant structure, but the 
[\sii]$\lambda$6717,6731/\ha, [\nii]$\lambda$6584/\ha, [\oiii]$\lambda$5007/\hb\ maps
 nicely depict the ionization structure of the region. They also obtain a reddening map whose 
 maximum correlates well with the maximum of the 24\,\mi\ emission, showing that the dust emitting 
 at this wavelength band is probably mixed with the ionized gas of the region and
 heated by the central ionizing stars. The authors also study the variation of several classical 
 emission line ratios proposed as metallicity tracers and find that while the R$_{\rm 23}$ 
index (Pagel et al. 1979) varies slightly within the region, [\nii]/\ha\ and [\nii]/[\oiii] 
show significant variations across the surface of NGC~595. 

Here, we present new photoionization models of NGC~595 in order to explain the results of the 3D
spectroscopic observations performed by Rela\~no et al. (2010) as well as the recently
observed spatial distribution of the IR emission. Due to its size and revealed shell structure,
NGC~595 is a very suitable object to apply photoionization models. It is 
the second most luminous \hii\ region in M33 and  presents an angular size of $\sim$1\arcmin, 
corresponding to a linear physical size of $\sim$250~pc \footnote{The adopted distance of M33 is 840~kpc 
(Freedman et al., 1991).}. The region has an \ha\ shell morphology that shows the action of the 
stellar winds of the massive stars located in its interior. Its age, 4.5$\pm$1.0~Myr,  was 
previously estimated by stellar photometry (Malumuth, Waller, \& Parker 1996) and by synthesis 
of integrated spectra in the far-ultraviolet (FUV) wavelength range (Pellerin 2006). 
The stellar content of the \hii\ region is made of $\sim$250 OB-type stars, $\sim$13 supergiants 
(Malumuth et al. 1996) and 9 spectroscopically confirmed Wolf-Rayet (WR) stars (Drissen
et al. 2008), plus a new WR star discovered using IFS (Rela\~no et al. 2010) and  located 
farther away from the other WR stars. The physical properties for NGC~595 were also derived 
using long-slit spectroscopic observations (V\'{\i}lchez et al. 1988) and echelle spectroscopy 
(Esteban et al. 2009). The electron temperature of the region is $\sim$8000~K, the electron 
density is consistent with the low density limit, and the estimated metallicity range is 12+log[O/H]=8.4-8.6, depending on the authors and the observational technique to derive it (V\'{\i}lchez et al. 1988; Esteban et al. 2009; Rela\~no et al. 2010). 

\begin{figure*}
\begin{minipage}{170mm}
\centerline{
\psfig{figure=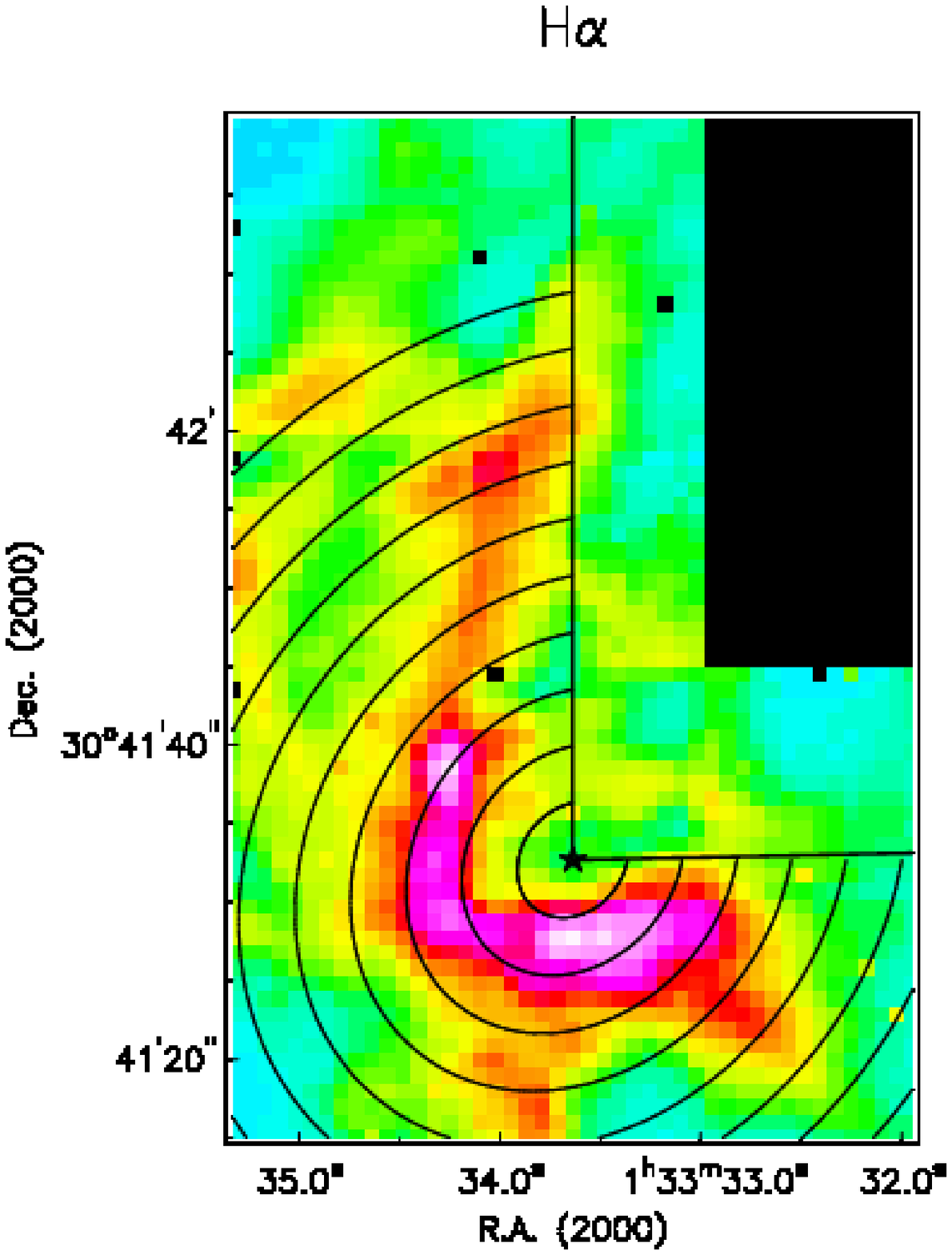,width=0.35\textwidth,clip=}
\psfig{figure=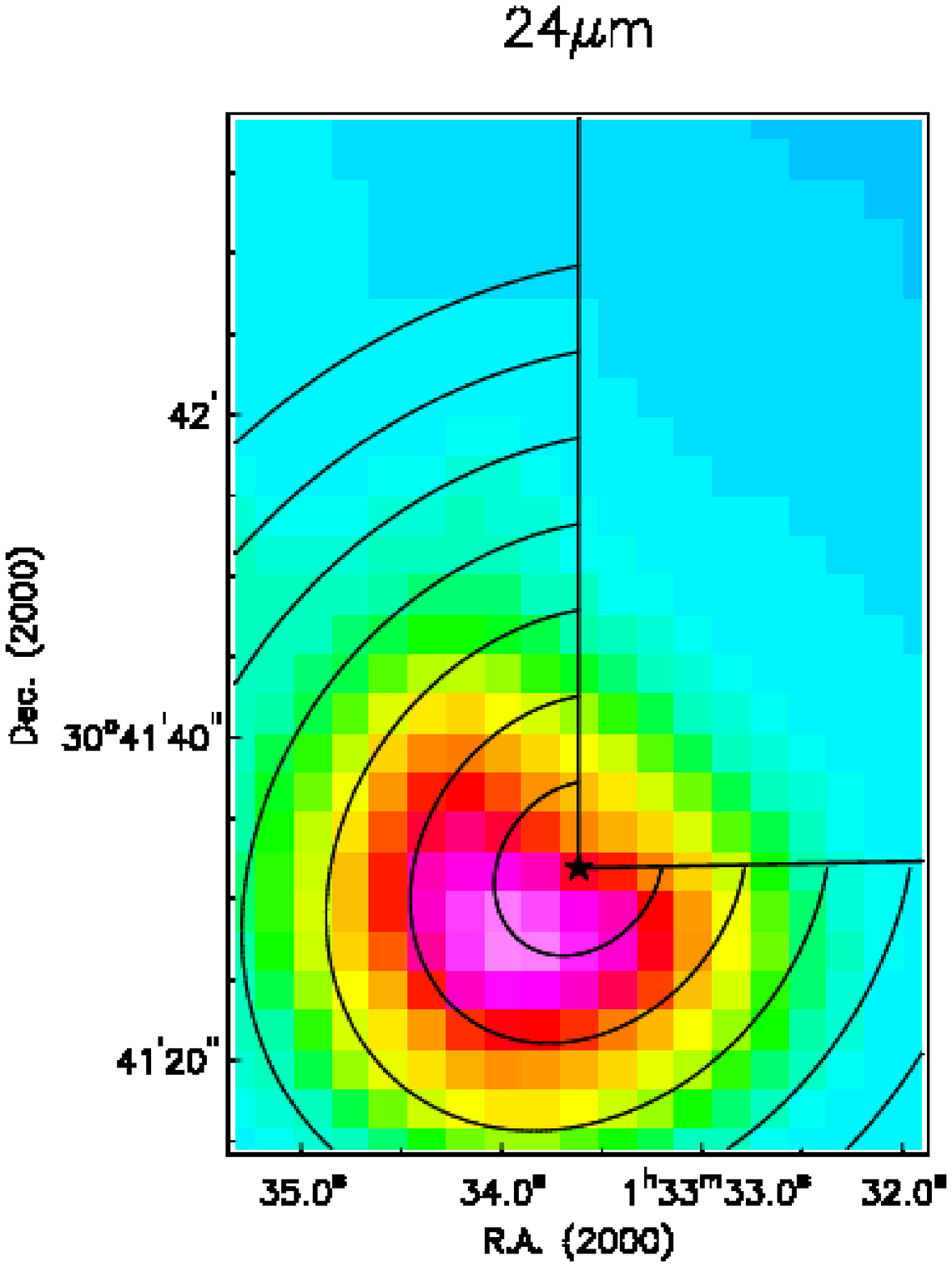,width=0.35\textwidth,clip=}
\psfig{figure=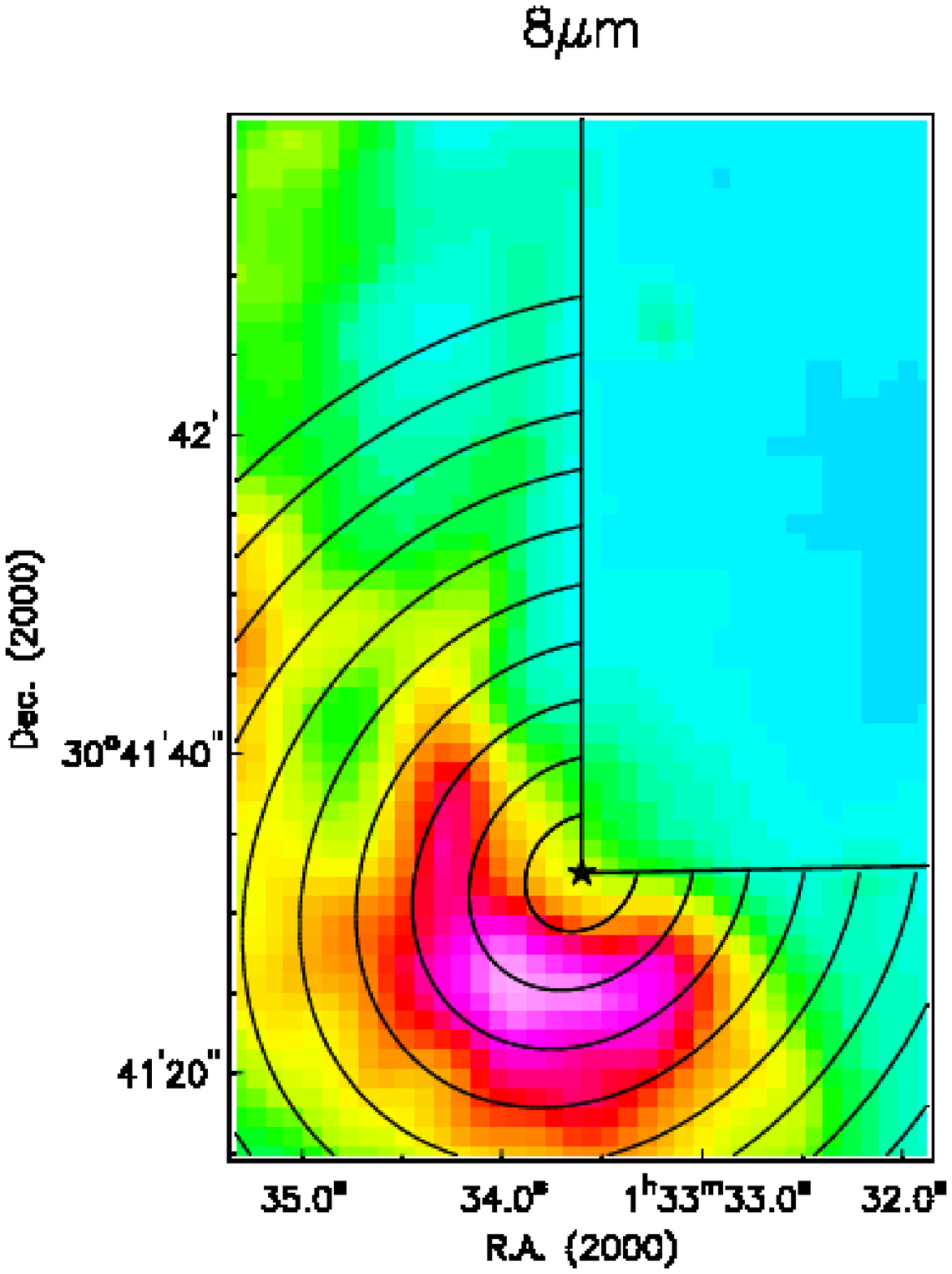,width=0.35\textwidth,clip=}}
\label{Fig1}
\caption{Left: map of the observed \ha\ flux derived from the IFS observations (Rela\~no et al. 2010). Center and right: 24\,\mi\ and 8\,\mi\ emission distributions, respectively. The elliptical annuli in the \ha\ and 8\,\mi\ images are separated by a distance of 4\,arcsec projected onto the major axis of the ellipse (a separation of 2\,arcsec was considered to obtain the integrated spectra for comparison to the models). In the case of the 24\,\mi\  the separation between annuli is 6\,arcsec. The black star shows the location of the main stellar clusters (see text).}
\end{minipage}
\end{figure*}


\section{Data sampling}

The optical observations we model here were taken at the 3.5~m telescope in
Calar Alto on the night of the 10th of October 2007. We used the lens array (LARR), made out of
16$\times$16 square elements, with the magnification scale of
1$\times$1\,arcsec$^2$. The small field of view of LARR in this configuration does not cover 
the total surface of the region, thus a mosaic of 13 tiles was needed in order to map the whole area. 
The V300 grating provides us with relatively low spectral resolution (3.40~\AA~pix$^{-1}$) but 
allows to cover the optical spectral range (3650-6990~\AA) we need for our modeling.  Further 
details of the observations and data reduction can be found in Rela\~no et al. (2010). 

The IR data of NGC~595 analyzed in this paper were taken from the Spitzer Data Archive: 
the 8\,\mi\ image from IRAC (Infrarred Array Camera, Fazio et al. 2004) and 24\,\mi\ - 170\,\mi\ from 
MIPS (Multiband Imaging Photometer, Rieke et al. 2004). The spatial resolutions of the images at 
8, 24, 70, and 160\,\mi\ are $\sim$2, $\sim$6, $\sim$18 and $\sim$40\,arcsec, respectively.
The stellar 
contribution of the 8\,\mi\ image was subtracted using the emission 
at 3.6\,\mi, following the method described in Helou et al. (2004) and Calzetti et al
 (2007). The IR observations and data reduction are explained in Rela\~no \& Kennicutt (2009) 
 for 8\,\mi\ and 24\,\mi, and in Verley et al. (2007) for the 70\,\mi\ and 160\,\mi. 

The power of the IFS observations is based on the coverage of the whole face of the region, which 
allows us to extract not only spectral but also spatial information of NGC~595. Given
 the \ha\ shell morphology of this particular \hii\ region, we decided to extract spectra 
 in elliptical concentric annuli covering the most intense zones of NGC~595 following this geometry. The spatial configuration of the elliptical 
 annulli are depicted in Figure~\ref{Fig1}. The ellipses are all centered at the location of the 
 main stellar clusters, (R.A. (J2000): 1h~33m~33.63s, DEC(J2000): 30d~41
m~32.6s), marked in the figure as a black star. The major to minor axis ratio of the ellipse 
is derived using the inclination angle of the galaxy (i=56\degree\ for M33, van~den Bergh,
2000), and we choose a position angle of 129\degree\ for the major axis since this orientation 
better traces the shell structure of the region (see Figure~\ref{Fig1}). In this configuration, 
we used rings of 2\,arcsec width to obtain the elliptical profiles from our IFS observations. 
We restrict our models to the most intense parts of the region emitting at \ha, which are covered by the ellipses depicted in Figure~\ref{Fig1}. 
The observed spatial configuration of this \hii\ region makes it possible to perform the analysis in this way. Moreover, the mid-IR emission distributions 
at 8\,\mi\ and 24\,\mi\ are concentrated in this part of the region, therefore using the zone marked in Figure~\ref{Fig1}, we completely take them into account when performing the modeling.

We generated masks to isolate the emission coming from each annulus. 
These masks were then applied to the data cube of our IFS observations and integrated spectra for each annulus 
were produced. The spectra, covering 3650-6990~\AA, were analyzed in the same way as 
explained in section~2.3 of Rela\~no et al. (2010). We used MPFITEXPR algorithm (Markwardt 2009) 
to fit the different emission lines with a Gaussian function plus a 1-degree polynomial function for the 
continuum subtraction. The final result of this procedure is a set of integrated fluxes for the emission lines
 fitted in the spectrum corresponding to each elliptical annulus. Flux errors were obtained as a combination of those derived in the profile fitting procedure and the uncertainty in the continuum subtraction (see P\'erez-Montero \& D\'\i az (2003) and references therein). Since the IFS observations were not taken under photometric conditions, no absolute calibration was performed, thus all the fluxes are in relative units. For the mid-IR bands, we use the same elliptical configuration as that used for IFS observations, but for the 24\,\mi\ observations we have chosen annuli of 6\,arcsec width, similar to the angular resolution of the 24\,\mi\ image. 
 

\begin{figure*}
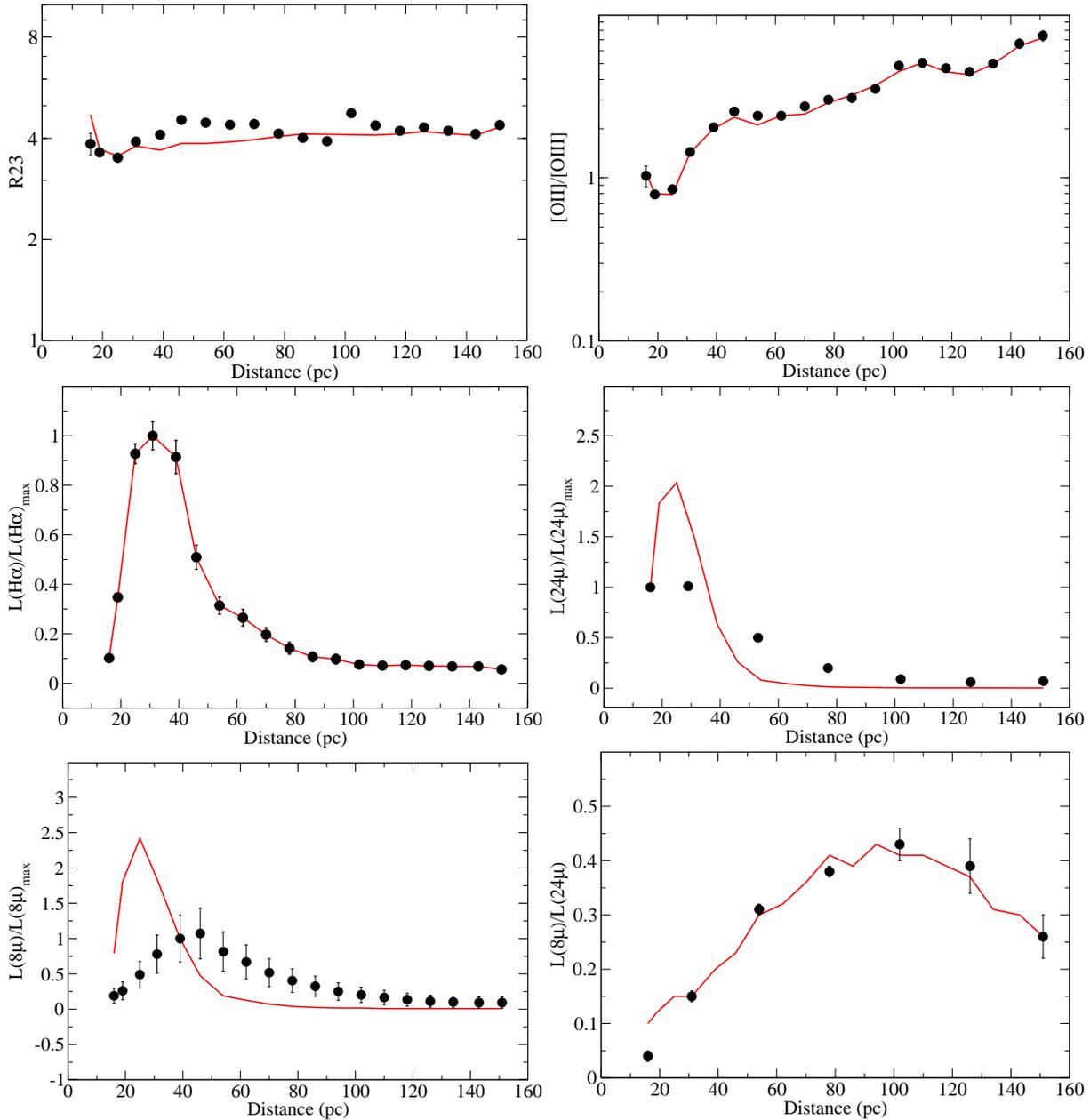

\begin{minipage}{170mm}
\centerline{
\psfig{figure=mn-n595models_fg02a.eps,width=8cm,clip=}
\psfig{figure=mn-n595models_fg02b.eps,width=8cm,clip=}}
\centerline{
\psfig{figure=mn-n595models_fg02c.eps,width=8cm,clip=}
\psfig{figure=mn-n595models_fg02d.eps,width=8cm,clip=}}
\centerline{
\psfig{figure=mn-n595models_fg02e.eps,width=8cm,clip=}
\psfig{figure=mn-n595models_fg02f.eps,width=8cm,clip=}}
\label{obs_mod}

\caption{Radial profiles of some physical properties in NGC~595 as observed (in black circles) and as derived from the
models (in red solid line).}
\end{minipage}
\end{figure*} 


From the integrated emission line fluxes for each annulus we can then derive other parameters normally used 
to study the properties of the gas. In the top panels of Figure~\ref{obs_mod} we show the \rdostres\ parameter (left panel) (\rdostres = ([\oii] $\lambda$3727 +  [\oiii] $\lambda\lambda$4959,5007)/\hb, Pagel et al. 1979) and the [\oii]/[\oiii] emission line ratio (right panel) as a function of the radial distance from the central stellar cluster.   [\oii]/[\oiii] changes over an order of magnitude: it shows low values closer to the location of the stars, indicating a high value of the ionization parameter, and rises towards larger distances from the stars where the ionization parameter declines. The \rdostres\ parameter, however, shows approximately the same value 
at any distance from the stellar cluster. This is indicative of the robustness of \rdostres\ to estimate the metallicity of 
\hii\ regions (see Rela\~no et al. 2010).  The \ha\ emission distribution normalized to the emission in the annulus with the largest
collected flux ({\em i.e.} the fourth one, situated at $\sim$7\,\arcsec) is shown in the left-middle panel of Figure~\ref{obs_mod}, revealing the \ha\ shell morphology of the region with a \ha\ maximum located at $\sim$30\,pc from the stellar clusters (see also left panel of Figure~\ref{Fig1}). The IR profiles are shown in the center-right panel for the 24\,\mi\ emission and in the lower-left panel for the 8\,\mi\ emission, normalized to the emission of the annuli 
with largest emission in each case. Both 8 and 24 \mi\ emission distributions show their maxima in a position outwards from the location of the stellar cluster (Figure~\ref{Fig1}). The lower-right panel shows the 8\,\mi/24\,\mi\ radial profile at the spatial resolution of the 24\,\mi\ map. This ratio can be useful to study the
nature of the dust, as 8\,\mi\  is expected to be mainly emitted by polycyclic aromatic hydrocarbons (PAHs) and it is then associated to
the photodissociation region (PDR) or the diffuse ionized gas; while the emission at 24\,\mi\ is mainly due to the contribution of small dust grains at relatively high temperature.
As we can see, there is an increase of the 8\,\mi/24\,\mi\ ratio from 0.04 in the first annulus up to a maximum of $\approx$ 0.4 in a
position much more outwards than the position of the maxima of the respective emissions in the 8\,\mi\ and 24\,\mi\ bands. Then it decreases slightly again.
The trend observed here for the 8\,\mi/24\,\mi\ shows that the hot dust emitting at 24\,\mi\ is partially mixed with the gas inside the \hii\ region while in the outskirts of the region the emission at 8\,\mi\ becomes more prominent, probably as the emission from the PAHs becomes stronger, then it
decreases again in the furthest annuli of the region.

\subsection{Dust Mass and extinction in NGC~595}
The mid-IR observations allow us to estimate the total amount of dust within the region and therefore obtain a measurement of the extinction that it produces. Then, we can compare the result with the extinction values obtained from the optical emission lines. We assume that most of dust mass is in the form of grains in thermal equilibrium, 
therefore the dust mass can be computed for a given flux (F$_\nu$) using the following expression:

\begin{equation}
\rm M_{dust}=F_{\nu}(T)D^{2}/\kappa_{\nu}B_{\nu}(T)
\label{eqdust}
\end{equation}
where $F_{\nu}(T)$ is the observed flux, $B_{\nu}(T)$ is the Planck function, $\kappa_{\nu}$ is the mass absorption coefficient ($\kappa_{100\mu m}$=63 cm$^2$g$^{-1}$, (Lisenfeld et al., 2002), with  $\kappa_{\nu}\sim\nu^{\beta}$, $\beta=2$, typical of interstellar grains (Draine \& Lee 1984) and D is the distance of NGC~595. To derive the dust mass we need to know the temperature of the dust, which is normally estimated from the F(70\mi)/F(160\mi) ratio (e.g. Tabatabaei et al. 2007). Using the 70\mi\ and 160\mi\ images of M33 from Spitzer (Verley et al. 2007) we integrated the emission in these two bands for NGC~595 and estimate a temperature for the dust of $\sim$30\,K. The result agrees with the values obtained by Tabatabaei et al. (2007) at the location of NGC~595 (see their fig.\,1) and is within the range of dust temperatures predicted recently by Kramer et al. (2010) for the central part of M33.  

\begin{figure}
\setcounter{figure}{2}
\psfig{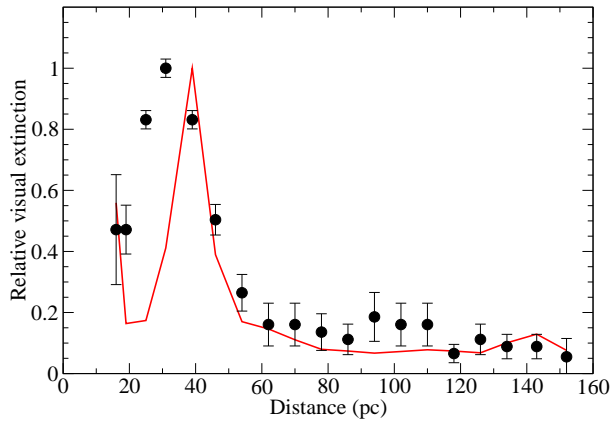}
\label{ext}
\caption{Radial profile of the observed (in black dots) and modelized (in red solid line) of the relative visual extinction.
The normalization value is the maximum of each radial distribution.}
\end{figure} 

Assuming a dust temperature of T=30\,K, we estimate the dust mass using the emission at 160\mi\ in Eq.\,\ref{eqdust}. The result is $\rm M_{dust}=1570\rm\,M_{\odot}$, which agrees with the dust mass range predicted by Viallefond, Donas \& Goss (1983). Taking into account the total gas mass of the region derived by Wilson \& Scoville (1992), 
$\rm M(HI)+M(H_{2})+M(H^{+})\sim$1.3$\times10^6\rm\,M_{\odot}$, we obtain a dust-to-gas ratio for NGC~595 of $\sim1.2\times10^{-3}$, a factor of $\sim$6 lower than the Galactic standard value ($\sim6.7\times10^{-3}$, Draine \& Lee 1984). Although this shows a significant discrepancy we note here that the metallicity of NGC~595 is $\sim$0.6 solar (Esteban et al. 2009), and therefore in principle we would expect a factor of two difference in the comparison of both dust-to-gas ratios. A possible reason for the relatively low dust-to-gas mass ratio might be the presence of cold dust not well traced by the emission at 160\,\mi. In the case we only take into account the mass of ionized gas (2.3$\times10^5\rm\,M_{\odot}$, from the same reference), the ratio increases up to $\sim6.8\times10^{-3}$.

We can also derive a measurement of the extinction that this amount of dust produces in the \hii\ region. 
Using the dust temperature estimated above and the integrated flux at 160\,\mi\ for the \hii\ region, we can predict the 
dust opacity at 160\,\mi, $\tau_{160}$$_{\mu}$$_m$, following the equation: 

\begin{equation}
\rm \tau_{160}=F_{160}(T)/\Omega\,B_{160}(T)
\label{eqext}
\end{equation}
\noindent where $\Omega$ is the solid angle covered by the region. 
Assuming $\tau_{160}$$_{\mu}$$_m$ $\simeq\,2200$\,$\tau_{H_{\alpha}}$ (Tabatabaei et al. 2007) we obtain an estimate of the extinction at \ha. 
We derive $\tau_{H_{\alpha}}$=0.18, equivalent to A(\ha)=1.086$\times\tau_{H_{\alpha}}$=0.19. 
This value agrees within the uncertainties with the intrinsic extinction derived using
the H$\alpha$/24 \mi\ ratio for NGC 595 (Rela\~no \& Kennicutt, 2009) and it is very 
close (a difference of 0.08 mag) to the extinction value derived from the 
Balmer decrement using PMAS data (Rela\~no et al. 2010).

However,  this extinction is not uniform across the nebula as it was already shown in Rela\~no et al. (2010). 
In Figure 3, we see in black dots the
radial profile of the relative visual extinction normalized to the maximum value. The observed extinction variation within the \hii\ region, which can be up to almost $\sim$1\,mag, agrees with the extinction variation ($\sim$1-1.5\,mag) found by Viallefond et al. (1983) using radio continuum emission. 

\begin{figure*}
\begin{minipage}{170mm}
\centerline{
\psfig{figure=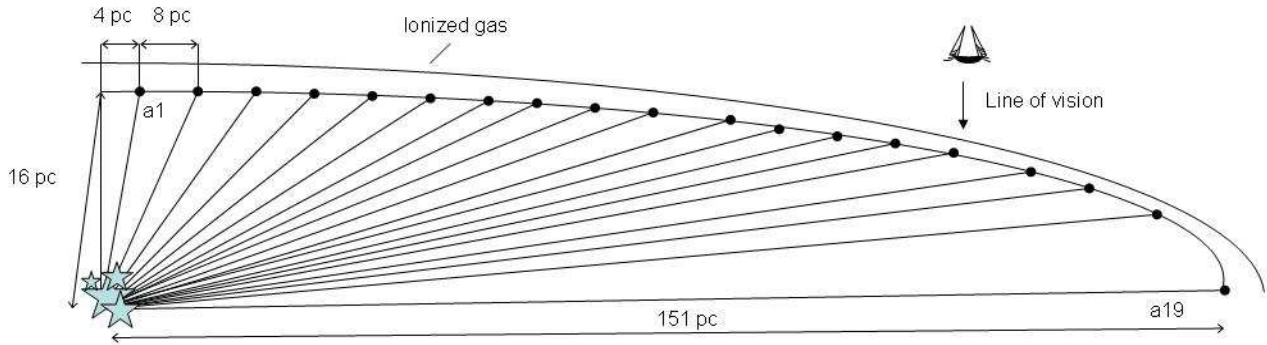,width=18cm,clip=}}
\label{mod_scheme}
\caption{Schematic representation of the adopted geometry of each of the 19 linear models made to reproduce the observed
features in each of the observed annuli of NGC~595. The line of vision indicates the orientation of the detectors to acquire the
available observational information taken for our models. The schematic representation here shows the dimension above 
the plane of the sky of the annuli plotted in Figure~\ref{Fig1}.}

\end{minipage}
\end{figure*}

\section{Model description}

Photoionization models were made to reproduce the observed properties in each 
annulus in both the optical and mid-infrared wavelength ranges. To do so, we resorted to
the photoionization code CLOUDY v.08c (Ferland et al. 1998).  
This code allows the modelization of the effects of a
radiation source situated in a determined spatial position on a one dimensional structure of gas and dust.
Then, it can be extrapolated to
a spherical close geometry or to an open geometry. Nevertheless,
the open geometry assumption can not be applied in this case as the collected IFS data
of NGC~595 do not provide the spatial distribution in
the direction perpendicular to the plane of projection. An alternative procedure
is to modelize each observed annulus with a single model of close geometry and to
compare both the observed and modelized luminosities per unit of area ({\em i.e.} surface brightness) relative to
the same luminosity in an arbitrary annulus.  
This procedure gives a qualitative rather than a quantitative analysis, 
but it constitutes a solid approach to study the variations of the physical
properties of the gas as a function of the distance to the ionizing source in each annulus.

We chose for our analysis the Spectral Synthesis Population (SSP) spectra from the
Starburst 99 libraries (Leitherer et al., 1999; V\'azquez \& Leitherer 2005), based on stellar atmosphere models from
Smith et al. (2002), Geneva evolutionary tracks with high stellar mass loss (Meynet et al., 2004), 
a Kroupa Initial Mass Function (IMF; Kroupa 2002) in two intervals (0.1-0.5 and 0.5-100 M$_\odot$)
with different exponents (1.3 and 2.3, respectively), the theoretical wind model (Leitherer et al. 1992)
 and a supernova cut-off of 8 M$_\odot$. The selection of this IMF is justified by the 
 detection of several supernova remants in the proximity of this HII region ({\em e.g.} Gordon et al., 1995).
The age of the ionizing burst was fixed to 4.5 Myr, according to the estimate made by Malumuth et al. (1996)
using HST/WFPC-2 ultraviolet and optical resolved photometry.
We fixed the metallicity of the stellar populations to Z = 0.02 (= Z$_\odot$),
which is the closer value to the oxygen abundance measured by Esteban et al. (2009).

The modelization of each annulus was made independent to each other,
assuming that each annulus has its own ionization structure originated
at different distances from the same ionizing cluster. 
This gives a total of 19 photoionization models, one for each annulus, as defined
in Section 2.
We assumed different initial input model conditions for each of the defined annuli and then we followed an iterative
method to fit the observed features, including the 
intensities of [O{\sc ii}] at 3727 {\AA} and [O{\sc iii}] at 4959 and 5007 {\AA} emission lines 
relative to H$\beta$ and the relative luminosity per unit of area of H$\alpha$ and 8\,\mi\ and 24\,\mi\
Spitzer bands.

We considered the same input gas-phase metallicities in all the annuli
as suggested by Rela\~no et al. (2010).
This was set to match the value derived by Esteban et al. (2009), including 
He, O, N, S, Ne, Ar and Fe elemental abundances derived from optical collisionally excited lines
and the measurement of the corresponding electron temperatures. The abundances for the rest
of elements were rescaled taking as 
reference the difference to the oxygen abundance in the solar photosphere as done by Asplund et al. (2005).
We assumed a constant density of 50 particles per cm$^3$, 
according to the values measured using the emission line ratio of [S{\sc ii}]
(and which range between 10 and 100 particles per cm$^{3}$ in the considered regions).

The inner radii of the ionized regions in each annulus can be initially estimated 
to reproduce the projected distances in the emission maps.
Hence, the projected distance between the first annulus and the ionizing source
is of 1'' (corresponding to 4 pc at the assumed M33 distance).
However, we considered a larger distance
for this closest annulus in order to optimize the fitting of the
observed quantities in the models. In this case, the best agreement between the observed
and modelized emission line ratios is found assuming a distance of 16 pc between the
inner face of the gas and the ionizing source. The difference between the distance found by
the models and the projected distance observed for this annulus
could be interpreted using a non-seen component of the distance in the direction perpendicular
to the plane of projection.
However, this deprojection is not required to find an agreeement between models and observations in
the furthest annuli. Therefore, the ratio between the adopted and projected distances is not maintained for all the other annuli, 
but it gradually decreases assuming the geometry of a hollow ellipsoid of revolution. 
Therefore, the considered factor of deprojection for the furthest annulus is 1
({\em i.e.} the distance measured in the projected image and the inner radius of
its corresponding model is the same, which is 151 pc in this region).
In Figure 4 we show
this geometry in a cut plane perpendicular to the plane of the projected sky.
This plot also shows the adopted
distances between the inner face of the ionized gas and the ionizing stellar cluster 
in the models of each annulus.

We left as free parameters the filling factor, the thickness of the gas shell and the amount of dust, which
vary in each iteration of the model in order to find the best agreement with the observed quantities.
In all models filling factor and thickness lead to a plane-parallel matter bounded geometry,
where a large fraction of the ionizing photons emitted by the central cluster escape to
the outer interstellar medium of the galaxy.
The dust-to-gas ratio  is a fundamental component to reproduce the ionization structure of the gas because
the dust heating affects its thermal balance.
We assumed the default grain properties of CLOUDY v08.00c, which has, essentially,  the properties 
of the interstellar medium and follows a MRN (Mathis, Rumpl \& Norsieck 1997) grain size distribution.

\begin{figure*}
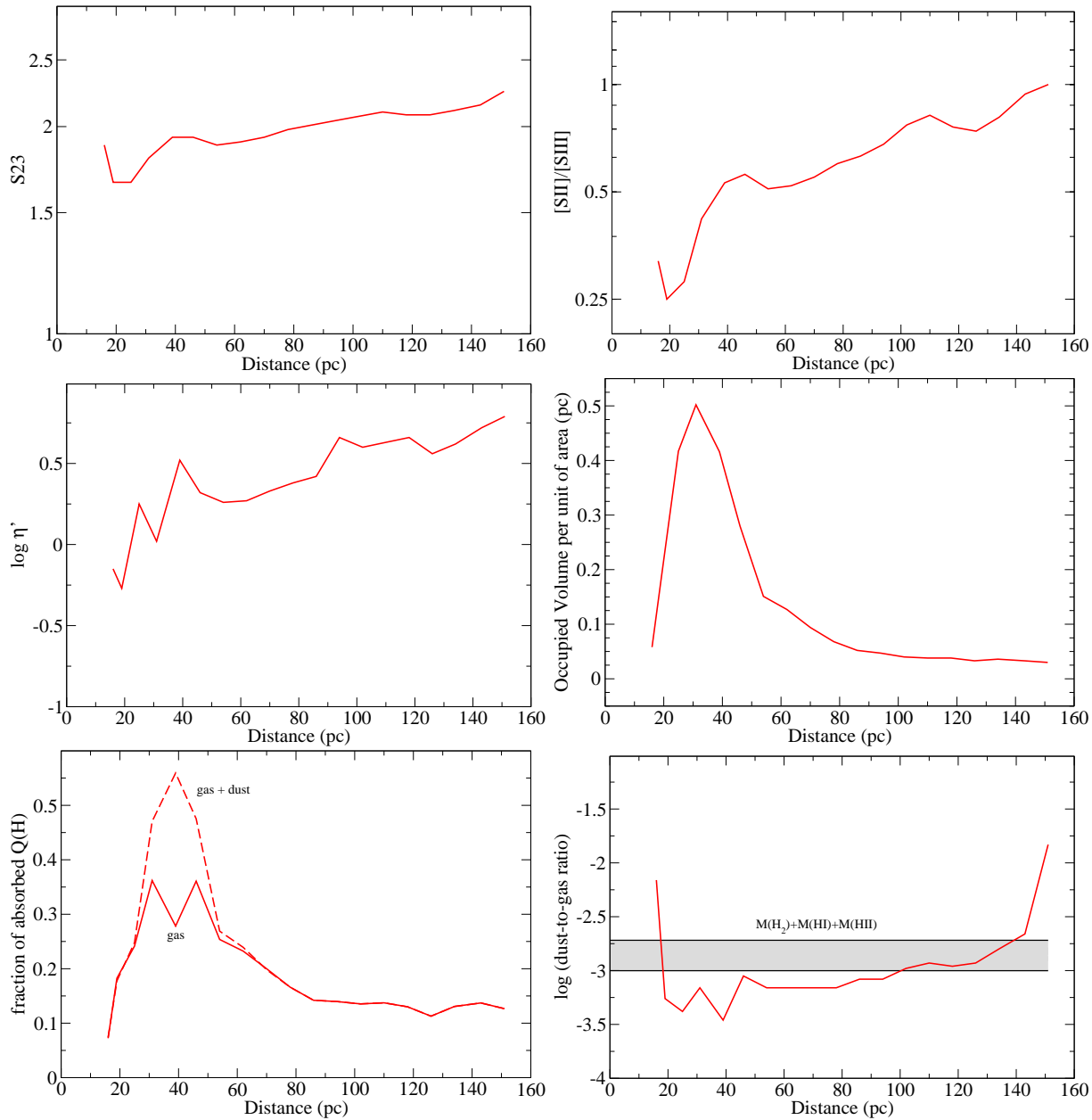

\begin{minipage}{170mm}

\centerline{
\psfig{figure=mn-n595models_fg05a.eps,width=8cm,clip=}
\psfig{figure=mn-n595models_fg05b.eps,width=8cm,clip=}}
\centerline{
\psfig{figure=mn-n595models_fg05c.eps,width=8cm,clip=}
\psfig{figure=mn-n595models_fg05d.eps,width=8cm,clip=}}
\centerline{
\psfig{figure=mn-n595models_fg05e.eps,width=8cm,clip=}
\psfig{figure=mn-n595models_fg05f.eps,width=8cm,clip=}}
\label{obs_mod2}
\caption{Radial profiles of some physical properties as predicted by the
models (in red solid line). See the text for details.}
\end{minipage}
\end{figure*} 

\section{Results and discussion}

In Figure~\ref{obs_mod} we show in solid red line the predictions made by the
model for each annulus in comparison with the same observed properties.
As these same properties were taken as input parameters in the
models, the agreement in the most part of them is excellent.
Hence, the R$_{23}$ parameter is well traced by the models, as a consequence
of the same adopted metallicity for all the considered annuli. 
Regarding the [O{\sc ii}]/[O{\sc iii}], this ratio is well reproduced across all the structure
as a consequence of the ionization parameter decrease, which varies in
the models from log U = -2.91 in annulus 1 to log U = -3.49 in annulus 19.
Other optical lines than [\oii] or [\oiii] can trace both the metallicity and ionization
parameter across the region, as it is the case of the sulphur emission lines [\sii] at
6717, 6731 \AA\ and [\siii] at 9069, 9532 \AA. 
In upper-left panel of Figure 5, we see the variation of the S$_{23}$
parameter, defined as the intensity sum of [\sii] and [\siii] relative to H$\beta$ (V\'{\i}lchez \& Esteban, 1996), 
which is sensitive to the metal content of the gas in a wide range of metallicity
(D\'\i az \& P\'erez-Montero, 2000). As we can see, the rate of variation 
is quite similar to that observed in R$_{23}$ ($\approx$ 20\%).
Regarding the ionization parameter
the ratio of these sulphur emission lines, [\sii]/[\siii], can also trace the decrease in the ionization
degree of the nebula (D\'\i az, 1988) following the same trend as [\oii]/[\oiii],
as we can see in the upper-right panel of Figure 5. This indicates
that sulphur lines supply an alternative in the red part of the optical spectrum to
oxygen emission lines in the blue to derive the ionization state and the metal content
of ionized gaseous nebulae.

Despite of the dependence of both, oxygen and sulphur, emission line ratios on the ionization parameter, as the
ionization potentials of the involved oxygen and sulphur ions are close,
the quotient between these two ratios cancels its dependence on ionization to first order
parameter, keeping a certain dependence on the shape of the ionizing 
stellar source ({\em i.e.} the equivalent effective temperature, Vilchez \& Pagel, 1988).
This quotient is defined as the $\eta'$ parameter:

\begin{equation}
\eta' = \frac{[OII]/[OIII]}{[SII]/[SIII]}
\end{equation}

\noindent which increases for lower stellar effective temperatures. The corresponding
$\eta$ parameter is also similar but it depends on the relative ionic abundance ratios.
For this case, as no chemical inhomogeneities were detected,
the two indicators are expected to behave in similar manners.
However, as we can see in middle-left panel of Figure 5, although we
used the same stellar synthetic atmosphere in all models, this parameter
increases from the inner to the outer annuli in almost an order of magnitude, 
showing a big jump of 0.75 dex in the central 40 pc ({\em i.e.} the location of the 
H$\alpha$ shell). 
This indicator, defined originally for integrated observations of giant \hii\ regions, must be
considered carefully in gas shells and matter-bounded geometry 
when studying the spatial variations of the effective temperature across a region with IFS data.

The relative surface brightness of H$\alpha$ in each model is also well reproduced,
with a maximum of this emission in the annulus 4, identical as derived from
the PMAS IFU observations. The two most important
input parameters in the models controlling this relative emission across the nebula are
the filling factor ($\epsilon$) and the thickness of the shell in each matter-bounded model.
As it is possible in models to increase the relative emission of the Balmer
emission lines by both increasing these two parameters, it is not possible
with these data to break the degeneracy of them and it is necessary to define a
 parameter which allows a direct comparison between the
input conditions of each model.
In this way, we can define the occupied volume ($V_o$), as:

\begin{equation}
V_o = \epsilon \times V(r_i,r_o)
\end{equation}

\noindent where $\epsilon$ is the filling factor and $V$ is the total volume of the gas in a thin shell, hence depending
on inner and outer radius of each annulus. In the middle-right panel of Figure 5 we
show the variation of the occupied volume for each annulus. The maximum in the 
relative emission of H$\alpha$ matches with a maximum in the occupied volume, which is in
turn due to a possible increase of the filling factor or the thickness of the nebula in
these positions.

Regarding the infrared, the emission of the nebula predicted by the models 
in the 8\,\mi\ and 24\,\mi\ Spitzer bands
were derived by subtracting the incident to the transmitted
continuum in each model to remove the stellar contribution.
Then we convolved the resulting spectrum with the shape of the corresponding Spitzer
filters.  As in the case of the H$\alpha$ emission, as models were calculated
in a close spherical geometry, only surface brightness were compared to the observations.
As it can be seen in the middle-right and lower-left panels of Figure~\ref{obs_mod}, 
the radial profiles of the relative emission of both 
8\,\mi\ and 24 \mi\ are reproduced by the models up to some extent, with a larger relative emission of both maxima
in the models. Besides, the position of the maximum in the 8\,\mi\ panel is predicted by the models in
a position slightly closer to the cluster than in the observations. 

Nevertheless, by varying
the dust-to-gas ratio in each model the 8\,\mi/24\,\mi\ ratio was  
perfectly fitted.  The measured values of this ratio across the nebula
range from 0.04 in the inner regions to a maximum of about 0.4 at 100 pc, but
it decreases again at further positions.  This decrease is explained by the
models by an enhancement of the dust-to-gas ratio in these positions.
The values of the 8\,\mi/24\,\mi\ ratio measured in NGC 595 are lower 
in all cases than the average values measured by 
Bendo et al. (2008) for a sample of nearby spiral galaxies. These authors
point out that in this sample the PAHs in the diffuse cold gas are expected to be responsible
for most of the 8\,\mi\ emission. As our models predict that NGC 595
is a matter-bounded shell with no PDR, no PAHs are expected to appear and
the emission at the 8\,\mi\ is emission mainly dominated by the IR continuum emitted by gas
and dust at this wavelength. On the other
hand, the 24\,\mi\ emission is high in this region due to the fraction of small hot dust grains 
surviving in the ionized gas.
These grains are also responsible for the absorption of a non-negligigle fraction
of the ionizing photons. In lower-left panel of Figure 5 we see
the fraction of these photons ({\em} i.e. with an energy larger than 13.6 eV)
absorbed both by the gas (in red solid line) and by the gas and the dust
(in dashed red line).  The most part of these photons escape from the gas
shell in those annuli with a lower occupied volume ({\em i.e.} with a
lower filling factor or a lower thickness), and the relative fraction of
absorbed photons by dust is almost negligible. However, in those regions
whose occupied volume increases to fit the relative emission of H$\alpha$, the
fraction of absorbed photons, which is more than a half here, is due
similarly to gas and dust. This is due to a larger fraction of bigger dust grains,
that survive at further distances, also enhancing the emission at 24\,\mi. 
This also affect the relative visual extinction due to the dust grains in each annulus.
In Figure 3, we see the comparison between the relative visual extinction derived from the
Balmer decrement in each annulus as compared to the values predicted by the models. 
The models fit fairly well the pattern of variation with a maximum of the extinction in
the same annulus of the fraction of occupied volume and, hence, the fraction
of absorbed ionizing photons is larger. This same region corresponds to the maximum in the
relative emission at 24\,\mi.

Finally, in the lower-right panel of Figure 5 we show the dust-to-gas ratio
obtained to fit the 8\,\mi/24\,\mi\ ratio. This is relatively constant  for the most part of
the annuli in a value about 10$^{-3.2}=6.3\times10^{-4}$, with the exception of the first annulus,
where it is of 10$^{-2}$ and in the the last two annuli, where it also increases.
These values are within the range of estimates made for the dust mass in the region based on the 
IR bands from Spitzer (see subsection 2.1).

\section{Summary and conclusions}

We carried out a set of photoionization models to describe the spatial distribution of the 
optical and mid-IR properties of
NGC~595, the second most luminous \hii\ region in the disk of the spiral galaxy M33. 
These models reproduce for the first time and simultaneously the spatial
radial profile of several observed properties in both the optical and the mid-IR.
We used the Integral Field Spectroscopy observations in the optical spectral range from
the PMAS instrument in the CAHA 3.5 m telescope (see Rela\~no et al. 2010) and the 
maps of 8\,\mi\ and 24\,\mi\ taken by the Spitzer telescope.
The input conditions extracted from the sets of spatially resolved
observations were divided in elliptical annular regions around
the central position of the \hii\ region, occupied by the ionizing
stellar cluster. Then, we derived the spatial profiles of the
H$\alpha$, 8\,\mi\ and 24\,\mi, relative surface brightness, the relative [\oii] and [\oiii] emission
line intensities and the extinction variation across the region. We also took as input
conditions of our models the gas metallicity reported by Esteban et al. (2009; 12+log(O/H) = 8.45)
and the age of the central stellar cluster, 4.5 Myr, derived by Malumuth et al. (1996).

This information was taken into account in the photoionization code CLOUDY
v. 08.00c (Ferland et al. 1998), assuming that each analyzed annulus can be
considered as an independent \hii\ structure.
The distance between the inner radius of the gas shell and the central
cluster was set to fit the projected measured distance of each annulus
but taking into account a certain deprojection factor in the case of the closest annuli.
Since no information about the distribution of the gas and the dust in the
axis perpendicular to the plane of projection can be collected, we 
considered spherical closed shells of gas in our models and then we
varied the filling factor, the shell thickness and the dust-to-gas
ratio in an iterative method in order to fit the corresponding surface brightness and relative variations
of the observed properties.

Our models reproduce fairly well the uniformity of the R$_{23}$ parameter, indicative of
an homogeneous metal content across the nebula, and the increase of the [\oii]/[\oiii] ratio,
as a consequence of the decrease of the ionization parameter. They also indicate that
the same parameters based on sulphur emission lines ({\em i.e.} [\sii] and [\siii]) are also
valid to trace these variations across the radial distribution of the \hii\ region.
Nevertheless a more accurate analysis of the spatial variation of the metal content of
this nebula could be carried out with the corresponding maps of the electron
temperature.
However, despite of a single common ionizing source, models indicate a $\sim$1 dex
variation of the $\eta'$ parameter (V\'\i lchez \& Pagel 1988), which is related to the 
equivalent effective temperature of the ionizing source.
This result indicates that this parameter must be used with care for the study of
the spatial variations of the ionizing field in \hii\ regions
which are matter-bounded or present a complex geometry.

Our models also reproduce the variation of the H$\alpha$ surface brightness, which
peaks at a distance of about 30 pc and decreases at further distances.
According to the models, this variation is due to the combination of a
matter-bounded geometry of the gas shell in all annuli and the variation of the occupied volume across
the shell. The occupied volume depends on the filling factor and the
thickness of the shell, but our models are not able to distinguish between these
two parameters to explain the enhancement in the H$\alpha$ emission.
According to these models, the number of ionizing photons escaping from the nebula varies between
a 45\% in the maximum of occupied volume and about a 90\% in the furthest
annuli.

Although the 8\,\mi\, is not well reproduced, the 8\,\mi/24\,\mi\ ratio fits fairly well with
the predictions of our models. This ratio, which varies between 0.04 at
the closest annulus and peaks at 0.4 in the outskirts of the nebula, is much lower
than the values measured by Bendo et al. (2008) in radial profiles of spiral galactic
disks. As models predict that all annuli are matter bounded and hence no 
photodissociation region is formed at close distances of the ionizing source, the 
PAHs dominating the 8\,\mi\, emission are much weaker than in the diffuse cold
gas regions of those galaxies. On the other hand, as 24\,\mi\, emission is 
mainly due to small hot grains surviving in the ionized gas, the 8\,\mi/24\,\mi\ ratio
is low in all the modelized annuli.

The geometrical variation of the occupied volume accross the region also shows 
some implications in the extinction structure of the region. Therefore, as the number
of hydrogen ionizing photons absorbed by the dust is only increased in
the regions with a larger occupied volume, the extinction also increases in
the same region, according to the PMAS observations.
Regarding the dust-to-gas ratio, models predict values compatible with
the estimates made using the gas mass from Wilson \& Scoville (1992) and
the dust mass derived from the integrated 70 and 160\,\mi\, bands from Spitzer. 
The dust-to-gas ratio estimated for the whole region ($\sim$1.2$\times10^ {-3}$) is
lower than in the Solar Neighbourhood, but consistent if we take into account 
that the metal content of this \hii\ region is subsolar.  In this sense, the corresponding
spatial distribution of the emission at 70\,\mi\ and 160\,\mi\ would supply the
spatial distribution of the dust-to-gas ratio to be directly compared with the
results from the models presented in this work but the spatial resolution up to now does not allow to perform this study. Future observations from Herschel telescope will be unique to address this issue.

\section*{Acknowledgements}
This work has been supported by the projects  AYA2007- 67965-C03-02 of the Spanish National Plan for Astronomy and Astrophysics
and CSD2006 00070 "1st Science with GTC" of the Spanish Ministry of Science and Innovation (MICINN).
This research was also supported by a Marie Curie Intra European Fellowship within the 7$^{\rm th}$ European Community Framework Programme.
We would like to thank Simon Verley for kindly providing us with the SPITZER 70\,\mi\ and 160\,\mi\ images, to David Mart\'\i n-Gord\'on for his
help with the convolution of synthetic spectra and Spitzer filters and to Ute Lisenfeld for very helpful discussions. 
This work is based in part on observations made with the Spitzer Space 
Telescope, which is operated by the Jet Propulsion Laboratory, 
California Institute of Technology under a contract with NASA.


\begin{thebibliography}{}
\bibitem[Bendo et al.(2008)]{2008MNRAS.389..629B} Bendo, G.~J., et al.\ 
2008, MNRAS, 389, 629 

\bibitem[]{} Calzetti, D., et al. 2007,  ApJ, 666, 870
\bibitem[]{} D{\'{\i}}az, A.~I., 1988, MNRAS, 231, 57
\bibitem[D{\'{\i}}az 
\& P{\'e}rez-Montero(2000)]{2000MNRAS.312..130D} D{\'{\i}}az, A.~I., \& P{\'e}rez-Montero, E.\ 2000, MNRAS, 312, 130 
\bibitem[]{} Draine, B. T. \& Lee, H. M. 1984, ApJ, 285, 89
\bibitem[]{} Drissen L.,  Crowther P.~A.,  {\'U}beda L., Martin, P. 2008, MNRAS, 389, 1033
\bibitem[]{} Esteban C.,  Bresolin F.,  Peimbert M., Garc\'{\i}a-Rojas, J., Peimbert, A., 
Mesa-Delgado, A. 2009, ApJ, 700, 654
\bibitem[]{} Ercolano B.,  Bastian N., Stasi{\'n}ska G.,  2007, MNRAS, 379, 945
\bibitem[]{} Fazio, G. G., et al. 2004, ApJS, 154, 10
\bibitem[]{} Freedman W.~L.,  Wilson C.~D.,    Madore B.~F.,  1991, ApJ, 372, 455
\bibitem[Gordon et al.(1998)]{1998ApJS..117...89G} Gordon, S.~M., Kirshner, 
R.~P., Long, K.~S., Blair, W.~P., Duric, N., 
\& Smith, R.~C.\ 1998, ApJS, 117, 89 

\bibitem[]{} Helou, G., et al. 2004, ApJS, 154, 253
\bibitem[]{} Kramer, C. et al. 2010, A\&A, astro-ph1005.2563
\bibitem[Le F{\`e}vre et al.(2003)]{2003SPIE.4841.1670L} Le F{\`e}vre, O., 
et al.\ 2003, SPIE, 4841, 1670 
\bibitem[]{} Lisenfeld, U.,  Israel, F. P., Stil, J. M., Sievers, A. 2002, A{\&}A, 382, 860
\bibitem[]{} Markwardt C.~B. 2009, PASP, 411, 251
\bibitem[]{} Malumuth E.~M.,  Waller W.~H., Parker J.~W.,  1996, AJ, 111, 1128
\bibitem[]{} Pagel B. E.~J.,  Edmunds M.~G.,  Blackwell D.~E., Chun, M. S., Smith, G. 1979, 
MNRAS, 189, 95
\bibitem[]{} Pellerin A.,  2006, AJ, 131, 849
\bibitem[P{\'e}rez-Montero 
\& D{\'{\i}}az(2003)]{2003MNRAS.346..105P} P{\'e}rez-Montero, E., \& D{\'{\i}}az, A.~I.\ 2003, MNRAS, 346, 105 
\bibitem[]{} Rela\~no, M. \& Kennicutt, R. C. Jr. 2009, ApJ, 699, 1125
\bibitem[]{} Rela\~no, M., Monreal-Ibero, A., V\'{\i}lchez, J. M., Kennicutt, R. C. 2010, MNRAS, 402, 1635
\bibitem[]{} Rieke, G. H., et al. 2004, ApJS, 154, 25
\bibitem[]{} Tabatabaei, F. S., Beck, R., Kr\"ugel, E., Krause, M., Berkhuijsen, E. M., Gordon, K. D., Menten, K. M. 2007, A{\&}A, 475, 133 
\bibitem[]{} S{\'a}nchez S.~F.,  Cardiel N.,  Verheijen M. A.~W., Mart\'{\i}n-Gord\'on, D., 
V\'{\i}lchez, J. M., Alves, J. 2007, A{\&}A,  465, 207
\bibitem[]{} van~den Bergh S. 2000, The galaxies of the Local Group, Cambridge University Pr
ess, Cambridge, UK
\bibitem[]{} Verley, S., Hunt, L. K., Corbelli, E., Giovanardi, C. 2007, A{\&}A,  476, 1161
\bibitem[Vilchez \& Esteban(1996)]{1996MNRAS.280..720V} Vilchez, J.~M., \& Esteban, C.\ 1996, MNRAS, 280, 720 
\bibitem[Vilchez \& Pagel(1988)]{1988MNRAS.231..257V} Vilchez, J.~M., \& Pagel, B.~E.~J.\ 1988, MNRAS, 231, 257 
\bibitem[]{} V\'{\i}lchez, J. M., Pagel, B. E. J., D\'{\i}az, A. I., Terlevich, E., Edmunds,
 M. G. 1988, MNRAS, 235, 633 
\bibitem[]{} Viallefond, F., Donas, J. \& Goss, M. 1983, A\&A, 119, 185  
 \bibitem[]{} Wilson, C. \& Scoville, N. 1992, ApJ, 385, 512
\bibitem[Wood et al.(2004)]{2004MNRAS.348.1337W} Wood, K., Mathis, J.~S., 
\& Ercolano, B.\ 2004, MNRAS, 348, 1337 

\end{thebibliography}
\end{document}